\documentclass[12pt,preprint]{aastex}
\usepackage{lscape}

\slugcomment{To appear in ApJ}
\shorttitle{X-rays from Binary CSPNe}
\shortauthors{Montez et al.}
\begin{document}
\title{X-ray Emission from the Binary Central Stars of the \\ Planetary Nebulae HFG 1, DS 1, and LoTr 5}
\author{Rodolfo Montez Jr.\altaffilmark{1}}
\author{Orsola De Marco\altaffilmark{2}}
\author{Joel H. Kastner\altaffilmark{1}}
\and 
\author{You-Hua Chu\altaffilmark{3}}
\altaffiltext{1}{Rochester Institute of Technology, Rochester, NY 14623, USA}
\altaffiltext{2}{Macquarie University, Sydney, NSW 2109, Australia}
\altaffiltext{3}{University of Illinois at Urbana-Champaign, Urbana, IL 61801, USA}
\begin{abstract} 
Close binary systems undergoing mass transfer or common envelope interactions can account for the morphological properties of some planetary nebulae.  
The search for close binary companions in planetary nebulae is hindered by the difficulty of detecting cool, late-type, main sequence companions in binary systems with hot pre-white dwarf primaries. 
However, models of binary PN progenitor systems predict that mass accretion or tidal interactions can induce rapid rotation in the companion, leading to X-ray-emitting coronae.
To test such models, we have searched for, and detected, X-ray emission from three binary central stars within planetary nebulae: the post-common envelope close binaries in HFG 1 and DS 1 consisting of O-type subdwarfs with late-type, main sequence companions, and the binary system in LoTr 5 consisting of O-type subdwarf and rapidly rotating, late-type giant companion. 
The X-ray emission in each case is best characterized by spectral models consisting of two optically-thin thermal plasma components with characteristic temperatures of $\sim 10$ MK and 15-40 MK, and total X-ray luminosities $\sim 10^{30}$ erg s$^{-1}$.
We consider the possible origin of the X-ray emission from these binary systems and conclude that the most likely origin is, in each case, a corona around the late-type companion, as predicted by models of interacting binaries.
\end{abstract}

\section{Introduction}

Close binary systems have been found in $17\pm5\%$ of the central stars of planetary nebulae (CSPNe) \citep{2009A&A...505..249M}. 
These systems are typically comprised of a hot, compact white dwarf (WD) or pre-WD primary and a low mass, late-type, main sequence companion and, hence, are likely the products of common envelope interactions \citep{1976IAUS...73...75P,1993PASP..105.1373I}.
In such systems, the primary star overfills its Roche lobe as it evolves onto the red giant branch (RGB) or asymptotic giant branch (AGB), engulfing the main sequence companion and forming a common envelope (CE).
Drag forces in the dense CE drain orbital energy from the binary, bringing the two stars closer together.  
The CE phase ends when either the companion spirals onto the primary's core; or the envelope is ejected.  
In the latter case, the residual is a close binary called a post-common envelope binary (PCEB).
Although PCEBs in PNe are thought to derive from CE interactions that took place during the AGB phase of the primary or secondary, other post-CE binaries may have originated in CE interactions on the RGB \citep[e.g.,][]{2003A&A...406..305S}.

The ejected CE can account for the morphological properties of some PNe.
\citet{1990ApJ...355..568B} argue that PNe with dense equatorial waists, often called butterfly PNe, form from a CE ejection immediately following the primary's early AGB phase, 
while elliptical PNe form from a CE ejection following a later AGB phase.
Recent observational \citep{2009A&A...505..249M} and hydrodynamical \citep{2008ApJ...672L..41R} studies appear to bear out these assertions. 
Indeed, within the PNe community, there is momentum building behind the idea that the entire diverse population of PNe may be the result of binary interactions \citep{2009PASP..121..316D}.

In most studies of PCEBs emphasis is placed on the evolution of the primary star into a mass accreting WD, i.e., a cataclysmic variable (CV). 
However, in this paper we are more interested in the evolution of the late-type, secondary companions.
The rapid rotation in young, single stars generates magnetic fields strong enough to power bright X-ray coronae \citep{1995A&A...300..775M}; however, as these stars age, their magnetic activity decays \citep[see reviews in][]{2004A&ARv..12...71G, 2009A&ARv..17..309G}.  
Hence, the old, late-type, companions in PCEB are not expected to exhibit strong coronal activity unless their rotation rates have increased by accretion of angular momentum during the CE phase \citep{1996MNRAS.279..180J, 2002ApJ...570..245S}. 
Although the degree to which the companion may undergo such accretion is uncertain \citep[e.g.,][]{1991ApJ...370..709H, 2008ApJ...672L..41R}, there is compelling evidence from extreme UV and X-ray observations of PCEBs that such a process occurs. 
From EUV spectra, \citet{2003ApJS..145..147S} find evidence for strong coronae around the cool companions in close binaries, while abundance anomalies determined from high resolution X-ray spectra of the close binary V471 Tauri show that the K dwarf companion must have accreted the highly processed and enriched material liberated by the primary on its RGB ascent \citep{2003ApJ...594L..55D}. 

By analogy with main sequence stars, coronae around the cool companions in PCEBs should be characterized by thermal plasmas from a few MK to tens of MK and therefore should emit hard X-rays with energies $\ga$ 0.5 keV.
\citet{2002ApJ...570..245S} argue that for CSPNe, a plasma temperature above 10 MK and $L_{X} > 5 \times 10^{29} \textrm{ ergs s}^{-1}$ would serve as compelling evidence for a "reborn" corona around a spun-up companion. 
There is little or no coronal contribution expected from the hot primary, since coronal activity from hot WDs or pre-WDs with effective temperatures $>$30 or $>$60 kK, respectively, is difficult to maintain when photospheric convection ceases \citep{1971A&A....12...21B, 1985A&A...152..107G}. 
Furthermore, if the CE is ejected, the magnetic field is ejected along with the CE \citep{2007MNRAS.376..599N}. 
Any remaining magnetic field will slow down the rotation of the remnant AGB core via magnetic braking \citep{2001Natur.409..485B}, resulting in decreased coronal activity. 
Indeed, \citet{2003AJ....125.2239O} discovered numerous sources of hard X-ray emission from WDs and argued that in all but two cases, this hard X-ray emission is a result of either accretion from a companion onto the WD or coronal activity on the companion itself.  

Thus, the potential exists to use X-ray observations as a tool for detecting otherwise unseen binary companions to CSPNe.
As an initial test, we targeted two binary systems, HFG 1 and DS 1, with the Chandra X-ray Observatory (CXO), and analyzed archival data from the serendipitous observations by CXO and the XMM-Newton X-ray Observatory (XMM) of the binary CSPNe in LoTr 5.  
All three CSPNe were detected in these X-ray observations.  
In this paper, we present analysis and interpretation of the X-ray emission from these three binary CSPNe.  
In Section 2, we summarize the properties of these binary systems; in Section 3 we describe the data and analyses; and in Section 4, we discuss the possible interpretations of the X-ray emission and their implications for models of binary interaction and PN shaping.

\section{Target CSPNe} 

The binary systems we consider are the PCEBs in the PNe \object[PN HFG 1]{HFG 1} and \object[PN DS 1]{DS 1}, and the evolved binary in \object[PN LoTr 5]{LoTr 5}. 
Throughout the text, we will generally refer to both the central star and PN by the PN name.   
In each case, the primary, and presumed origin of the CE and PN, is a hot ($T_{eff} > 60$ kK) O-type subdwarf (sdO) evolving towards a WD, while the secondaries are cool, late-type stars (see Table~\ref{binarytable}). 
These systems are the precursors to cataclysmic variables (CV), but do not exhibit characteristics of active accretion, and they feature strong reflection effects, i.e., irradiation of the secondary by the primary. 

\subsection{HFG 1 and DS 1}

Properties of the PCEB systems in the PNe HFG 1 (V664 Cas) and DS 1 (KV Vel, LSS 2018) have been summarized by \citet{2008AJ....136..323D} and are listed in Table~\ref{binarytable}. 
Here, we briefly highlight the key properties. 
The binary orbital periods are well-determined due to strong reflection effects \citep[e.g.][]{2005MNRAS.359..315E,1996MNRAS.279.1380H}.
The binary period of V664 Cas (within HFG 1) is 14 hours \citep{1987BAAS...19..643G}.  
Assuming a primary mass $\sim 0.6M_{\sun}$, \citet{2005MNRAS.359..315E} determined a range of acceptable fits for the secondary mass ($M_2 \sim 0.4-1.1M_{\sun}$).
The spectral analysis by \citet{2004ARep...48..563S} suggests the chemical composition of the secondary in V664 Cas is near solar, but with some anomalies (overabundances of nitrogen, magnesium, and silicon) that are interpreted as enriched primary material accreted by the secondary during the CE phase. 
Distance estimates to HFG 1 by \citet{2005MNRAS.359..315E} range from 0.31 to 0.95 kpc, which is consistent with the statistical determination by \citet{1982A&A...114..414H} of 0.4 kpc and a distance of 0.6 kpc found from the PN surface brightness relation \citep{2008FrewThesis}.  
Here, we adopt a value of 0.6 kpc.

The binary period of KV Vel (within DS 1) is 8.5 hours \citep{1985ApJ...294L.107D}.   
\citet{1996MNRAS.279.1380H} determine $M_1 = 0.63M_{\sun}$ and $M_2 = 0.23M_{\sun}$ for the components of KV Vel.  
\citet{1984ApJ...278..702S} used the color excess measured from KV Vel to estimate a distance range of 0.5 to 1.4 kpc. 
We adopt the distance of 0.7 kpc obtained via the PN surface brightness relation by \citet{2008FrewThesis}.

The PNe HFG 1 and DS 1 are large and faint \citep{1990ApJ...355..568B}.  
HFG 1 shows strong evidence of interacting with the interstellar medium \citep{1982A&A...114..414H,1985A&A...143..475H} and \citet{2009MNRAS.396.1186B} recently discovered a trail of shocked material behind the star as it plows through its local environment. 
The PN DS 1 suggests a late elliptical PN with blown out edges \citep{1990ApJ...355..568B}, and \citet{2009A&A...505..249M} point out the presence of low ionization structures that suggest an outflow along the E-W axis of the PN.

\subsection{LoTr 5}

The binary system in the PN LoTr 5 (IN Com, HD 112313) is comprised of an sdO primary and a rapidly rotating G5III-IV companion \citep{1983ApJ...269..592F, 1997A&A...322..511S}.
\citet{1996A&A...307..200J} considered the similarities with rapidly rotating FK Comae stars \citep{1981ApJ...247L.131B}, arguing that FK Comae systems end the CE phase as a coalesced star, whereas LoTr 5 ended the CE phase as a close binary after ejecting its envelope.
As the binary orbital period is unknown, \citet{1996A&A...307..200J} consider a few alternatives: a) a very short period of $\sim 1$ day, b) a moderately short period of a few days, and c) a wide binary with a period of a few years.
The radius of the giant star ($\sim 12 R_{\sun}$) rules out the close separation required for a very short period binary, but beyond this lower bound, there is little constraint on the binary orbital period.          
The observed photometric variation of 5.9 days is attributed to the rotation of the giant companion \citep[see detailed discussion on the history of orbital solutions for LoTr 5 in][]{1997A&A...322..511S}.
This interpretation is supported by the presence of Ca II H \& K emission lines that are indicative of chromospheric activity due to rotation, and which imply a projected rotational velocity of $v\textrm{ sin} i = 67\textrm{ km s}^{-1}$ \citep{1997A&A...322..511S}.
 
Long slit spectra and narrow band images of the large ($\sim 6\farcm5 \times 1\farcm7$), faint PN are modeled by \citet{2004MNRAS.347.1370G} as a bipolar nebula inclined 17$^{\circ}$ to the line of sight.  
If the nebula is formed from the ejection of a CE, then the orbital plane is likely to coincide with the bipolar axis \citep[e.g.][]{1990ApJ...355..568B}.
Such a low binary inclination would then explain the inability to observe radial velocity variability from the binary system. 
However, \citet{1997A&A...322..511S} found that the inclination of the rapidly rotating giant is $i \sim 45^{\circ}$.  
If one takes the orbital inclination to be identical to that of the bipolar lobes, then this suggests that the binary and its components are not coplanar.  This is considered in \citet{1996A&A...307..200J} as evidence for a longer orbital period, since the components are not close enough to become synchronized via tidal interactions.  
\citet{1997A&A...320..913T} argue that the high projected equatorial velocity suggests LoTr 5 is a wide binary system, in which the rapid rotation inferred for the giant companion is due to accretion from the AGB wind and does not require a CE phase. 

Distance estimates to LoTr 5 are widely discrepant, ranging from 0.5 to 6.9 kpc \citep{2004MNRAS.347.1370G}; the closer distance is favored \citep[e.g.][]{1997A&A...322..511S}. 
We adopt a distance of $0.5 \textrm{ kpc}$, after \citet{2008FrewThesis}.
The high galactic latitude ($+88\fdg46$) of this PN renders interstellar reddening negligible. 

\section{Data and Analysis}

\subsection{Observations} 

We acquired targeted observations of HFG 1 and DS 1 on the back-illuminated S3 chip of the ACIS detector array onboard CXO. 
A summary of the CXO observations of HFG 1 and DS 1 is presented in Table~\ref{obstable}.  
In these two on-axis CXO observations we detect X-ray point sources centered on the positions of the central stars in HFG 1 and DS 1. 
There is no evidence for diffuse X-ray emission from either nebula (see also Section 4).
These CXO observations were prepared and analyzed according to data analysis threads accompanying the Chandra X-ray Center CIAO software (version 4.1) \citep{2006SPIE.6270E..60F}.
We find no high-background periods during the CXO observations of HFG 1 and DS 1. 

Archival data of the serendipitous observations of LoTr 5 from both CXO and XMM were obtained through the High Energy Astrophysics Science Archive Research Center, a service of the Astrophysics Science Division at NASA's Goddard Space Flight Center and the High Energy Astrophysics Division of the Smithsonian Astrophysical Observatory. 
A summary of the XMM and CXO observations of LoTr 5 is presented in Table~\ref{obstable}.
LoTr 5 lies $7\farcm3$ off axis in the XMM observation (ObsID 0012850201; 2002 June 06) of the galaxy group NSCS J125606+255746 and is detected on all three European Photon Imaging Camera (EPIC) detector arrays (pn, MOS1, and MOS2).  
The XMM observations were operated in Full-Frame Mode with the thin filter.  
Data were reprocessed completely from the Observation Data Files using the XMM-Newton Science Analysis Software (SAS) package version 7.1.0 with the calibration files available in Current Calibration File Release 241 (XMM-CCF-REL-241).
In the XMM observations of LoTr 5, the detected emission is consistent with a point source at the off-axis position of the CSPN.  
There is no evidence of diffuse X-ray emission associated with the nebula (see also Section 4).
The Chandra observation (ObsID 3212; 2002 December 04) of the same galaxy group field was obtained six months after the XMM-Newton observation. 
In this CXO observation, LoTr 5 lies $\sim8\farcm6$ off axis in the CXO observation, just within the field of view of the Advanced CCD Imaging Spectrometer (ACIS) detector array, on a CCD (the front-illuminated S4) that lies adjacent to the prime imaging CCD used to target the galaxy cluster (the back-illuminated S3).
Since the ACIS-S chip array is optimally aligned for use with the curved focal plane Rowland circle for grating spectroscopy, the S4 image of LoTr 5 is very out of focus.  Hence, it is difficult to ascertain the nature of the emission; however, its spectrum is useful.
The CXO data obtained for LoTr 5 were prepared and analyzed according to data analysis threads accompanying the CIAO software.
We filtered the XMM observations of LoTr 5 for high background periods and bad events using the standard filters for the imaging mode observations. 
No high background periods were found during the CXO observation of LoTr 5.
A summary of the observations of LoTr 5 is presented in Table~\ref{obstable}. 

\subsection{Spectral Analysis}

The source and background spectra from the on-axis X-ray point sources in HFG 1 and DS 1 were extracted using the \textit{psextract} task in CIAO.  
The source and background spectra from the off-axis, serendipitous XMM and CXO sources at the position of the central star of LoTr 5 were extracted from source regions determined by the 90\% encircled energy radius for a point source at the appropriate source off-axis angles.
For the XMM observations of LoTr 5 the 90\% encircled energy radius is 45$^{\prime\prime}$ and for the CXO observation the 90\% encircled energy radius is $10^{\prime\prime}$. 
For each source spectrum, we generated target- and observation-specific response matrix (RMF) and ancillary response (ARF) files.
This extraction of the source and background spectra and responses accounts for the broadening and vignetting due to the off-axis angle of the serendipitous observations of LoTr 5. 

All spectral fits were performed with XSPEC version 12.3.1x \citep{1996ASPC..101...17A}.
We use reduced $\chi^{2}_{\nu}$ statistics to distinguish best fit models, where $\nu$ represents the number of degrees of freedom.  
When multiple models give acceptable $\chi^{2}_{\nu}$ values ($\chi^{2}_{\nu} < 2$) we use goodness-of-fit tests (\textit{goodness} in XSPEC), to determine the more appropriate model.
This goodness-of-fit test performs Monte-Carlo simulations of the spectrum as drawn from the best-fit model, and calculates the $\chi^{2}_{\nu}$ statistic for each simulation.  
A value of $\sim$50\% is returned if the data is accurately described by the set of simulated spectra, while values closer to 100\% suggest the data are not well described by the model. 
We performed 500 simulations per test.
Due to the small number of counts detected for HFG 1 and DS 1, we used Churazov weighting \citep{1996ApJ...471..673C} on the unbinned spectra.  
Such weighting is preferred for sources with low count rates and many empty spectral channels. 
Weighting is assigned with respect to the average of the surrounding channels.  
The best-fit spectral parameters (and their 90\% confidence levels) and the resulting inferred X-ray emission properties are presented in Table~\ref{specfits}.  
We now discuss specific aspects of the spectral fitting for each source.

\subsubsection{X-ray Spectrum of HFG 1}

\begin{figure}
\includegraphics[scale=1.0]{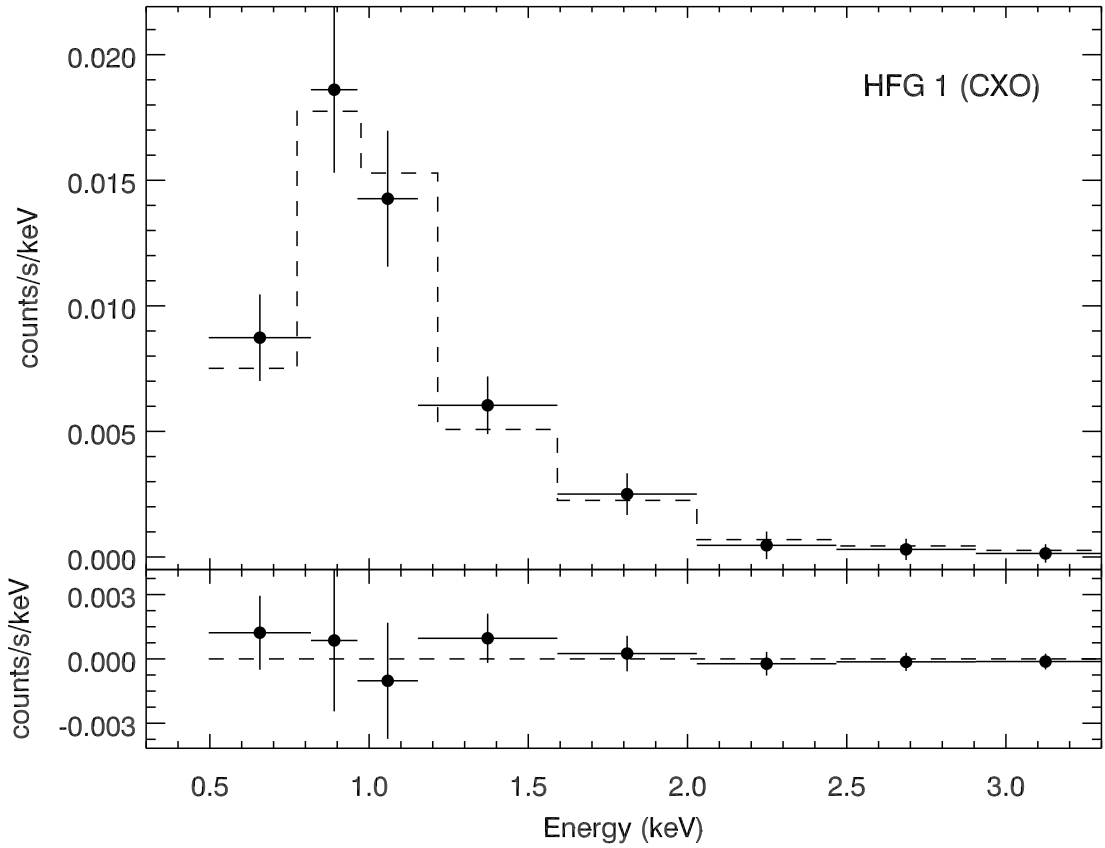}
\caption{\label{hfg1spec} X-ray spectrum of HFG 1 observed by CXO overlaid with the best-fit model (top) and residuals of the spectral modeling (bottom).
The best fit model (dashed line) requires two thermal plasma components at 6 MK and 21 MK, with intervening absorption of $N_H \sim 2.4 \times 10^{21}$ cm$^{-2}$.}
\end{figure}

We attempted to fit single-temperature-component (1-T) and two-temperature-component (2-T) optically thin thermal plasma (\textit{mekal}) models to the X-ray spectrum of HFG 1. 
Attempts to determine the intervening absorption from the model fits failed to constrain the absorbing column density ($N_H$), so we adopted an intervening absorption,  $N_{H} = 2.4 \times 10^{21} \textrm{ cm}^{-2}$, corresponding to the value of $E_{B-V}\sim0.5$ mag estimated to the central star \citep{2005MNRAS.359..315E}.
The absorbed 1-T model can only fit the data when the plasma abundance is allowed to vary.  
The best-fit 1-T model gives $\chi^{2}_{511} \sim 0.80$ and requires an abundance of 0.06 times solar \citep[with solar values given by ][]{1989GeCoA..53..197A}, with 90\% confidence range of 0.02 to 0.14 solar.
However, this abundance result is inconsistent with the conclusions drawn by \citet{2004ARep...48..563S} who found stellar abundances near or slightly above solar.  
The goodness-of-fit test yields 98\% of the simulations with a better $\chi^{2}_{\nu}$ value, suggesting the flexibility introduced by freeing the model abundance parameter simply allows the model to fit features that are due to noise.
Adding a second thermal plasma component improves the overall appearance of the fit and produces a similar $\chi^{2}_{\nu}$ value of $\chi^{2}_{510} \sim 0.79$.
The goodness-of-fit test yields 68\% of the simulations with a better $\chi^{2}_{\nu}$ value, however, indicating that the 2-T plasma model is a better description of the data than the 1-T, variable abundance model. 
The parameters of this best-fit 2-T model are presented in Table~\ref{specfits} and the model is overlaid on the source X-ray spectrum in Figure~\ref{hfg1spec}.

\subsubsection{X-ray Spectrum of DS 1}

\begin{figure}
\includegraphics[scale=1.0]{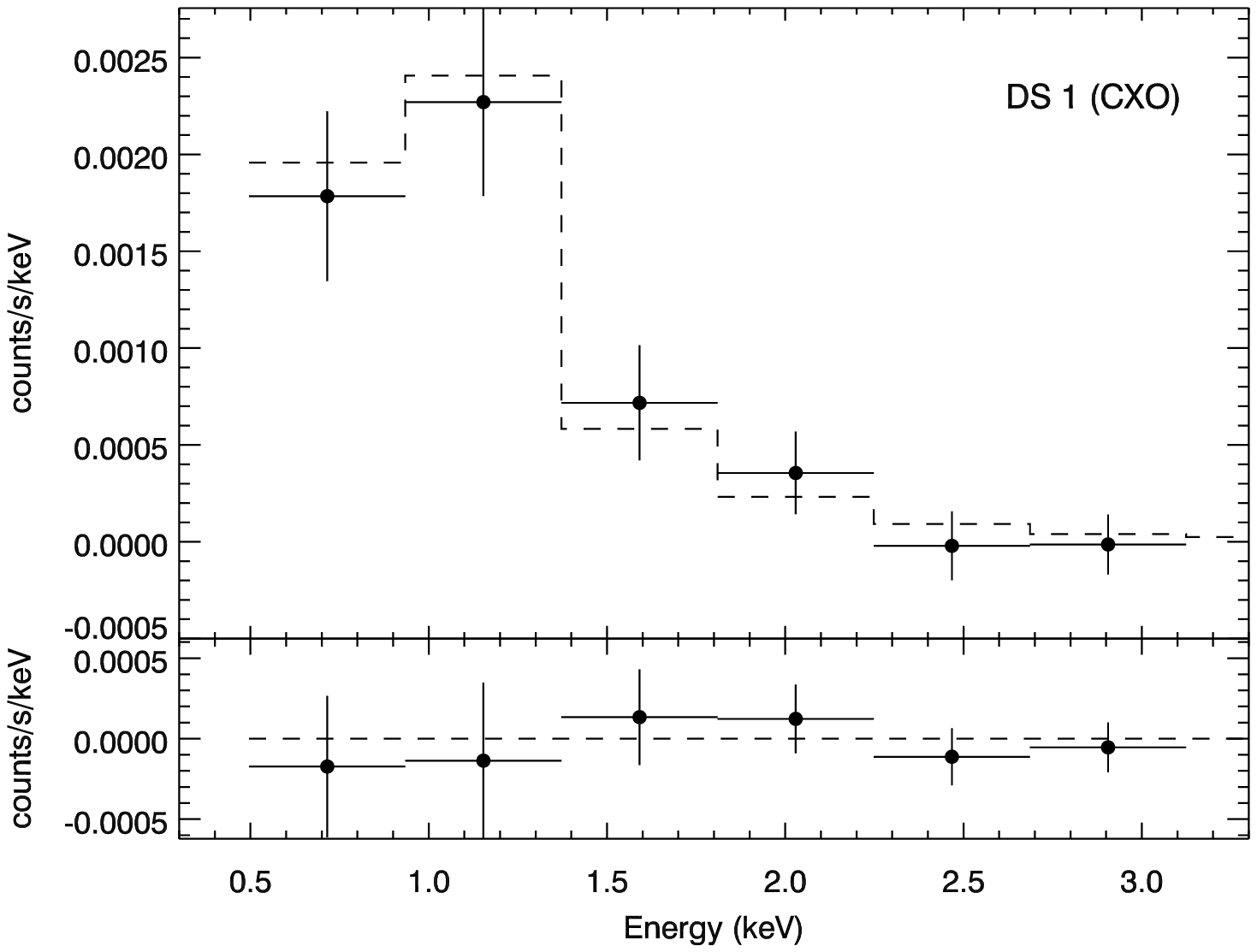}
\caption{\label{ds1spec} X-ray spectrum of DS 1 observed by CXO overlaid with the best-fit model (top) and residuals of the spectral modeling (bottom).  
The best fit model (dashed line) requires two thermal plasma components at 3 MK and 15 MK, with intervening absorption of $N_H \sim 8 \times 10^{20}$ cm$^{-2}$.}
\end{figure}

As in the case of HFG 1, the intervening absorption to DS 1 could not be constrained.  
We adopted $N_{H} = 8.3 \times 10^{20} \textrm{ cm}^{-2}$ based on the value of $E_{B-V}\sim0.15$ mag estimated toward the central star \citep{1984ApJ...278..702S}.
The best-fit 1-T model with solar abundances gives $\chi^{2}_{512} \sim 0.49$ for a hot component at $T_{X} \sim 14.3$ MK, while the best-fit 2-T model gives $\chi^{2}_{510} \sim 0.49$ for a similar hot component at $T_{X} \sim 14.5$ MK and a cooler component at $T_{X} \sim 3$ MK. 
The best-fit 1-T and 2-T models are indistinguishable, based solely on their $\chi^{2}_{\nu}$ values. 
The X-ray fluxes determined from the best-fit 1-T and 2-T models agree within their 90\% confidence ranges.  
We find from visual inspection, however, that the 2-T model locally improves the fit at the soft end of the spectrum.  
Hence, we adopt the 2-T model as the best fit.  
Additionally, the goodness-of-fit tests for the two models shows a modest improvement, from 85\% to 63\%, when the second component is added. 
Adopting the 2-T model over the 1-T model does not effect the overall interpretation of the X-ray emission observed from the central star of DS 1.
The parameters of the best-fit 2-T model are presented in Table~\ref{specfits}, and the model is overlaid on the source X-ray spectrum in Figure~\ref{ds1spec}.

\subsubsection{X-ray Spectra of LoTr 5}

\begin{figure}
\includegraphics[scale=1.0]{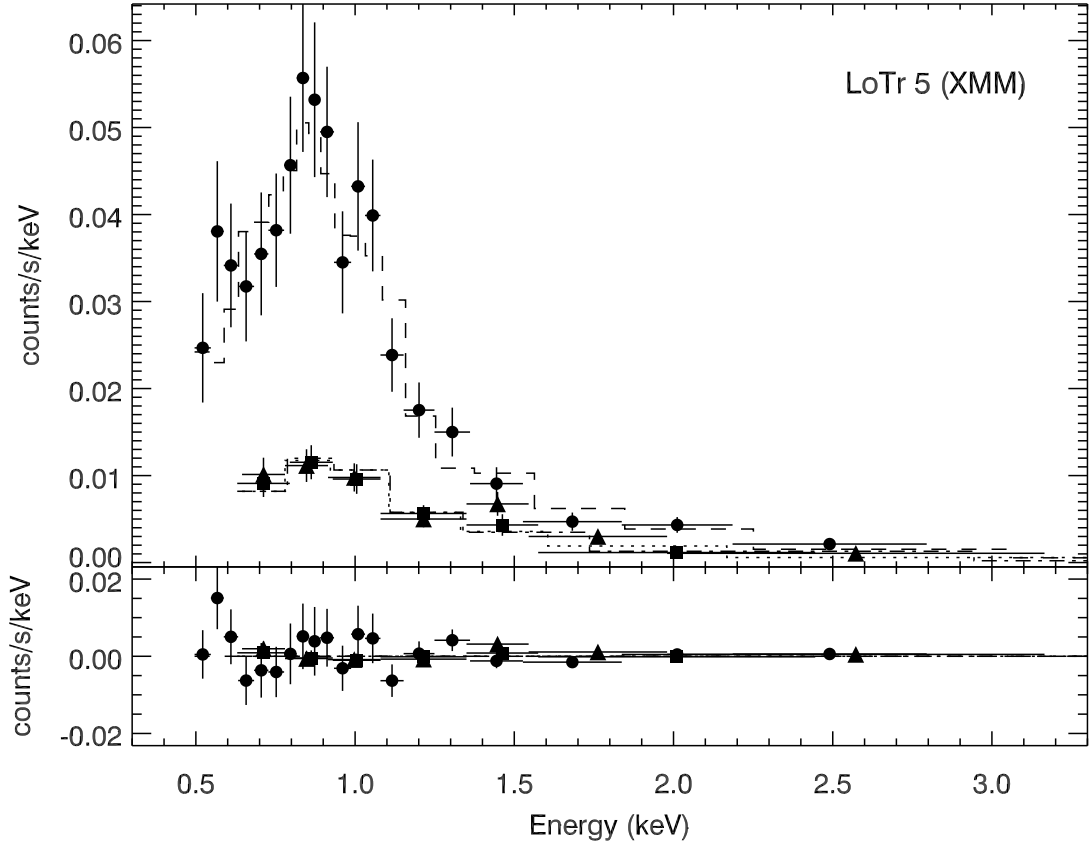} 
\caption{\label{lotr5xmmspec} X-ray spectra of LoTr 5 observed by the three EPIC detectors (pn - circles, MOS1 - triangles, MOS2 - squares) overlaid with the best-fit model (top) and residuals of the simultaneous spectral modeling (bottom).   
The best fit model (broken lines) requires two thermal plasma components at 7.6 MK and 26 MK.  No intervening absorption component is required, reflecting the small distance and high galactic latitude to LoTr 5. 
 }
\end{figure}

\begin{figure}
\includegraphics[scale=1.0]{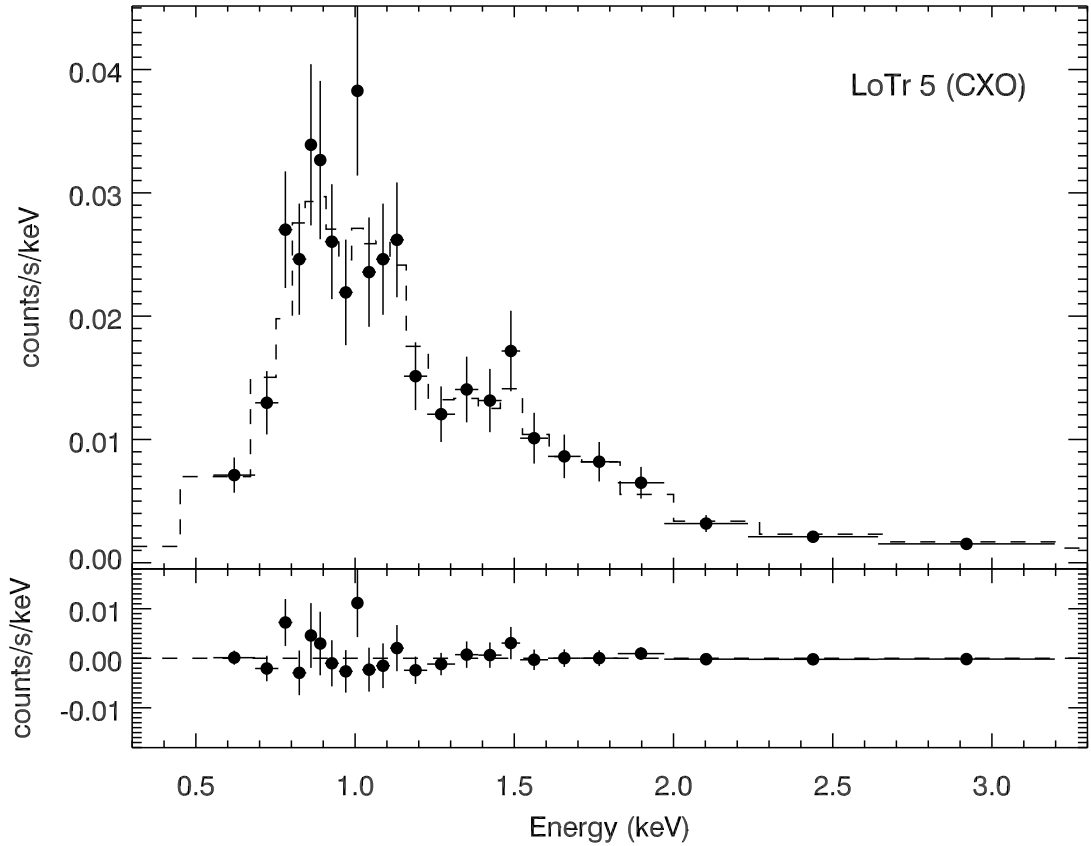}
\caption{\label{lotr5cxospec} X-ray spectrum of LoTr 5 observed by CXO overlaid with the best-fit model (top) and residuals of the spectral modeling (bottom).
The best fit model (dashed line) requires two thermal plasma components at 7.5 MK and 40 MK, with no intervening absorption.}
\end{figure}

\begin{figure}
\includegraphics[scale=1.0]{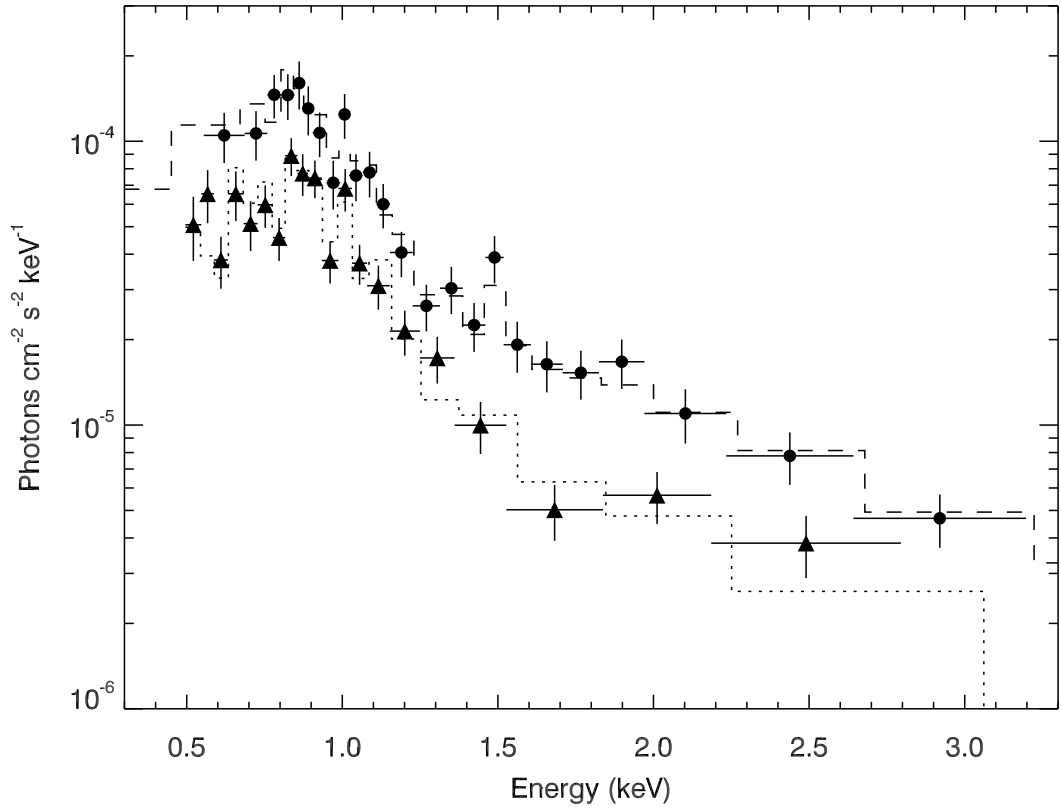}
\caption{\label{lotr5ufspec} Unfolded X-ray spectra of LoTr 5 observed by XMM EPIC pn (triangles) and CXO ACIS-S (circles) overlaid with the best-fit, two temperature plasma models (CXO - dashed line, XMM - dotted line).  
An increase in flux during the six month interval between the XMM and CXO observations is apparent across the energy range.}
\end{figure}

We attempted to fit single-temperature models with wide ranges of absorption and elemental abundances to the XMM and CXO spectra of LoTr 5, but all such fits resulted in $\chi^{2}_{\sim 30} \sim 3-6$.
The best-fit models of the X-ray spectra are instead comprised of two, optically thin, thermal plasma components.
Attempts to add foreground absorption to the model did not appreciably improve the fit and did not constrain the absorbing column density.  
This is consistent with the high galactic latitude and proximity of LoTr 5, which suggest there is negligible intervening absorption.
The parameters of the best fits of the X-ray spectra using 2-T models are presented in Table~\ref{specfits} and overlaid on the source X-ray spectra in Figures~\ref{lotr5xmmspec}~\&~\ref{lotr5cxospec}.
The difference between the apparent flux levels measured from the XMM and CXO observations is larger than the cross-calibration discrepancy between EPIC observations and CXO ACIS-S3 observations \citep{2002astro.ph..3311S}, however, there are no specific cross-calibration data comparing XMM EPIC observations and CXO ACIS-S4, where the X-ray emission from LoTr 5 is detected. 
To facilitate direct comparison, we present the unfolded spectra from the CXO ACIS-S and XMM EPIC pn detectors together in Figure~\ref{lotr5ufspec}. 
In the absence of any significant cross-calibration discrepancies, it appears that in the six month interval between the XMM and CXO observations the brightness increased across the entire energy range, indicating that the two spectral components arise from a common physical process. 

\subsection{Timing Analysis}

To test for variability, we performed Kuiper tests \citep{1992nrca.book.....P} to assess whether the cumulative distribution of the barycenter-corrected source and background photon arrival times can be described by a Poisson process. 
These tests are independent of temporal bin size, but do not provide information on the period or amplitude of variability. 
Where the Kuiper test suggested deviation from a Poisson process, we constructed background-subtracted light curves using for multiple temporal bin widths.   
These background-subtracted light curves were extracted using the same source and background regions used for spectral extraction (using the \textit{ltc1} option of the CIAO tool \textit{dmextract}). 
The low source count rates and relatively short temporal windows do not permit us to effectively use Fourier period detection methods or create phase-binned spectra. 
Hence, we restricted our assessments of source variability to the Kuiper test, comparision with the background, and visual inspection of the light curves. 

\subsubsection{HFG 1 and DS 1} 

\begin{figure}
\includegraphics[scale=1.0]{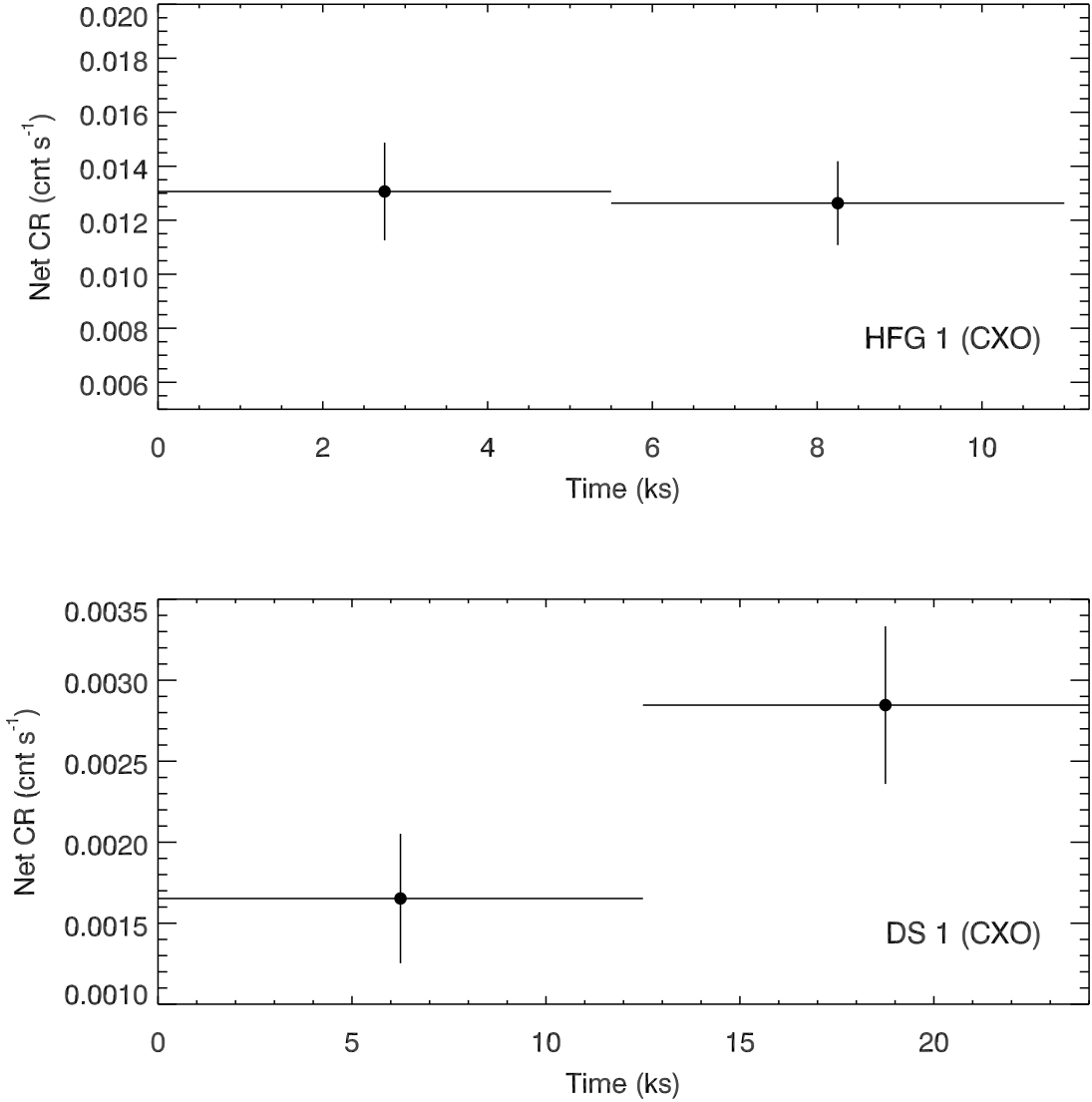}
\caption{\label{binarylc} Background-subtracted X-ray light curves of HFG 1 (upper panel) and DS 1 (lower panel).  The light curve of HFG 1 is consistent with a steady count rate, while DS 1 exhibits a marginally significant increase in the count rate during the second half of the observation.}
\end{figure}

The Kuiper test gives a 98\% and 87\% probability, respectively, that the photon arrival times of HFG 1 and DS 1 can be ascribed to a constant Poisson process.           
The minimum temporal bin widths to obtain a signal-to-noise ratio of 3 are 670 s and 3300 s for HFG 1 and DS 1, respectively. 
Below this signal-to-noise ratio the source count rate approaches the background rate and it becomes difficult to discern source variability from noise fluctuations.
At these temporal bin widths, the rms noise is on par with that expected from a Poisson distribution.   
Improving the signal-to-noise ratios by widening the light curve bins to half the exposure times for HFG 1 and DS 1 confirms the Kuiper test result that HFG 1 produces a steady count rate, and suggests that DS 1 may exhibit variability (Figure~\ref{binarylc}).     

\subsubsection{LoTr 5}

For LoTr 5, Kuiper tests yield only 16\% and 2\% probabilities, respectively, that the X-ray sources detected in the XMM and CXO observations can be described by a constant Poisson process.  
In both observations, this deviation between the source and a simulated constant Poisson process is not present in the background.
The minimum temporal bin width for a signal-to-noise ratio of 3 is 200 s and 320 s for the XMM and CXO observations, respectively. 
In the XMM observation, for temporal bin widths greater than 1000 s, the rms noise of the source is consistently 1.3 times that expected from a Poisson distribution. 
In the CXO observation, the rms noise of the source increases to 1.8 times that expected from a Poisson distribution. 
Since there are too few counts to construct phase-binned spectra, we cannot make proper inter-observatory comparisons of fluxed light curves.  
The background-subtracted light curves are presented in Figure~\ref{lotr5lc}.  
These light curves and the results of the Kuiper tests suggest that the X-ray source in LoTr 5 is modestly variable.
   
\section{Discussion}

As noted in Section 1, we anticipate that the X-ray emission from the central stars of DS 1, HFG 1, and LoTr 5 is due to the presence of coronal activity associated with the late-type companions in these likely or candidate PCEBs. 
However, there are other mechanisms which may give rise to thermal plasma emission in the range of temperatures we observe.  
We consider these alternative interpretations before discussing the coronal interpretation. 

\subsection{X-ray Emission from the WD}

White dwarfs can be sources of X-ray emission, if they have low opacities or 
high effective temperatures \citep[e.g.][]{1994A&A...290..834J,1997MNRAS.286..369M,1993A&A...268..561M}. 
However, WD photospheric emission is much softer than the X-ray emission detected from the three CSPNe considered in this paper.  
Theoretically, cool WDs with convective envelopes may generate magnetic fields and coronae and emit X-rays \citep{1990SvA....34..291S,1995ApJ...453..403T}, however, no such emission has been detected from field WDs \citep{2007ApJ...657.1026W} and the primaries in the three CSPNe considered here are too hot to sustain coronal activity.
Hard X-ray emission from single WDs at $T_{X}$ similar to those of the three CSPNe considered here is very rare; only two cases are known: KPD 0005+5106 and the central star of the Helix Nebula \citep{2003AJ....125.2239O,2001ApJ...553L..55G,2004AJ....127..477C}.   
It is possible, but statistically unlikely, that the X-ray emission detected from one or all of the three CSPNe considered here is of the same (unknown) origin. 

\subsection{X-rays from Colliding Wind Shocks}

Shocks from colliding winds associated with these CSPNe may come in two forms: (1) a fast stellar wind colliding with a slower, previously ejected, stellar wind---as in the interacting stellar wind (ISW) theory that successfully explains the emissivity patterns of PNe \citep{1978ApJ...219L.125K}; and (2) collision of the two winds generated by the components of low-mass binary systems.
Both types of energetic wind interactions can create X-ray emitting shocks with plasma temperatures in the range 1-8 MK and with X-ray luminosities from $10^{31}$ to $10^{33} \textrm{ erg s}^{-1}$ \citep[e.g.][]{1997A&A...319..201M,2008ApJ...672..957K}.
A significant contrast between the two types of colliding winds is the scale at which each occurs.
The ISW collisions form large hot bubbles that fill the PN cavity on the order of a few thousand AU \citep[e.g.][]{2000ApJ...545L..57K,2005ApJ...635..381M}, while colliding-binary winds occur in a thin, conical interaction region with a vertex located at the stagnation point of the two winds, typically on the order of the binary separation \citep{1995A&A...297L..87M}.

In the targeted, on-axis, CXO observations of HFG 1 and DS 1, and the off-axis XMM observation of LoTr 5,
there is no evidence for any extended X-ray emission.
In the case of NGC 40, the X-ray emission is of such a low surface brightness that it was only revealed via blind spectral extraction from the expected emission region \citep{2005ApJ...635..381M}.
We performed similar blind spectral extractions for the X-ray observations of the large PNe in LoTr 5, HFG 1, and DS 1 and found no evidence for diffuse, low surface brightness X-ray emission.
Furthermore, the X-ray luminosities determined from the X-ray spectral fits to the point source emission in these three, large PNe (Table~\ref{specfits}) are fainter than the faintest diffuse X-ray emitting plasma in PNe ($\sim 10^{31}$ erg s$^{-1}$) and the plasma temperatures are hotter \citep[see][]{2008ApJ...672..957K}.

The emitting volumes of the unresolved X-ray emission can be explored through the emission measures determined from the X-ray spectral fits.
The emission measure ($EM = \int n_e n_H dV \sim n_e^2 V$) of large, diffuse, X-ray emitting plasmas in PNe suggest
plasma densities of a few 100 cm$^{-3}$ \citep{2008A&A...489..173S}, whereas the emission measure from colliding binary winds
suggest densities of $1-5\times 10^{6} \textrm{ cm}^{-3}$ \citep{1995A&A...297L..87M}.  
For a typical emitting volume characteristic of colliding binary winds ($\sim 10^{41}$ cm$^{3}$)
the emission measure determined from the X-ray spectral fits (Table~\ref{specfits}) suggest
the density is $\sim 10^{6}$ cm$^{-3}$. 
This density is similar to that inferred for colliding binary winds in symbiotic stars \citep{1997A&A...319..201M}.
However, the X-ray luminosities reported in Table~\ref{specfits} are one to two orders of magnitudes fainter than such systems.
Additionally, in HFG 1 and DS 1, there is no evidence for a wind fast enough ($>500$ km s$^{-1}$) to account for colliding-binary wind shocks. 
\citet{1993ApJ...415..258M} reported a fast wind speed of 3300 km s$^{-1}$ from their analysis of the IUE high resolution UV spectrum of LoTr 5 that potentially could explain the temperatures determined from the X-ray spectra of LoTr 5, although the evidence for such a fast wind is disputed (Guerrero; private communication).

\subsection{X-rays from Accretion-related Processes}

Although there is no evidence for accretion disks in prior studies of HFG 1 and DS 1, given that these systems are candidate pre-cataclysmic variables, we should consider whether the X-rays from the core of each PN could be due to a hot accretion disk or from accretion onto the compact sdO primary. 
From \citet{2002apa..book.....F}, the maximum temperature for an accretion disk around a compact object (in this case, a white dwarf) is given by 
\begin{equation}
T_{max} = 120 kK \left[ \left(\frac{M}{0.6M_{\sun}}\right) \left(\frac{\dot{M}}{1\times10^{-7}M_{\sun}\textrm{ year}^{-1}}\right) \left(\frac{R}{0.01R_{\sun}}\right)^{-3} \right]^{1/4}.
\end{equation}
Considering the primary star masses and radii listed in Table~\ref{binarytable}, the maximum disk temperature for these three binary systems would exist around the compact primary in LoTr 5. 
However, this temperature, $T_{max} \sim 40$ kK, is much too cool to account for the detected X-ray emission, hence, the X-ray emission cannot arise from a hot accretion disk, but this does not preclude the existence of a disk, nor a disk as a source of accretion material. 

If material is falling onto the compact object from Roche lobe overflow (RLO) from the companion, or from the inner region of an accretion disk.
Such material can reach velocities up to the free fall velocity, $v_{ff} = (GM/R)^{1/2}$, forming shocks in the boundary layer (BL) between the flow and compact object.
From \citet{2002apa..book.....F}, the temperature of the shocked material is given by
\begin{equation}
T_{ff} = 150 MK \left(\frac{\mu}{0.6}\right)\left(\frac{M}{0.6M_{\sun}}\right)\left(\frac{R}{0.01R_{\sun}}\right)^{-1}. 
\end{equation}
The resulting implied temperature for the compact companion in LoTr 5 would be $T_{ff} \sim 90$ MK.  This is higher than the temperature determined from the X-ray spectral fits, though not high enough to confidently dismiss as the origin of the X-ray emission. 
Meanwhile, the implied temperature for the compact companions in HFG 1 and DS 1 is in the range $T_{ff} \sim  8-10$ MK, which is consistent with the range we determine from the X-ray spectral fits. 

Since an accretion disk and/or infalling material can imply the presence of RLO, we consider the possibility of RLO in the three binary CSPNe.
We estimate the Roche lobe ($R_L$) radius using the approximation given in \citet{1983ApJ...268..368E}, 
\begin{equation} 
\frac{R_L}{A} \approx \frac{0.49 q^{2/3}}{0.6 q^{2/3} + ln(1+q^{1/3})}, 
\end{equation}
where $A$ is the orbital semi-major axis and $q = M_1/M_2$. 
We find that a few of the possible solutions given by \citet{2005MNRAS.359..315E} for the binary system parameters in HFG 1 suggest that the main sequence companion can fill its Roche lobe, but the solution provided by \citet{2004ARep...48..563S} suggests it cannot (see Table~\ref{binarytable}). 
The binary system parameters determined by \citet{1996MNRAS.279.1380H} suggest that the main sequence companion in DS 1 is unable to fill its Roche lobe and, therefore, is not capable of actively supplying material for accretion streams or accretion disks. 
The lack of support for RLO suggests that accretion is likely not the source of the observed X-ray emission from HFG 1 and DS 1.
 
\begin{figure}
\includegraphics[scale=1.0]{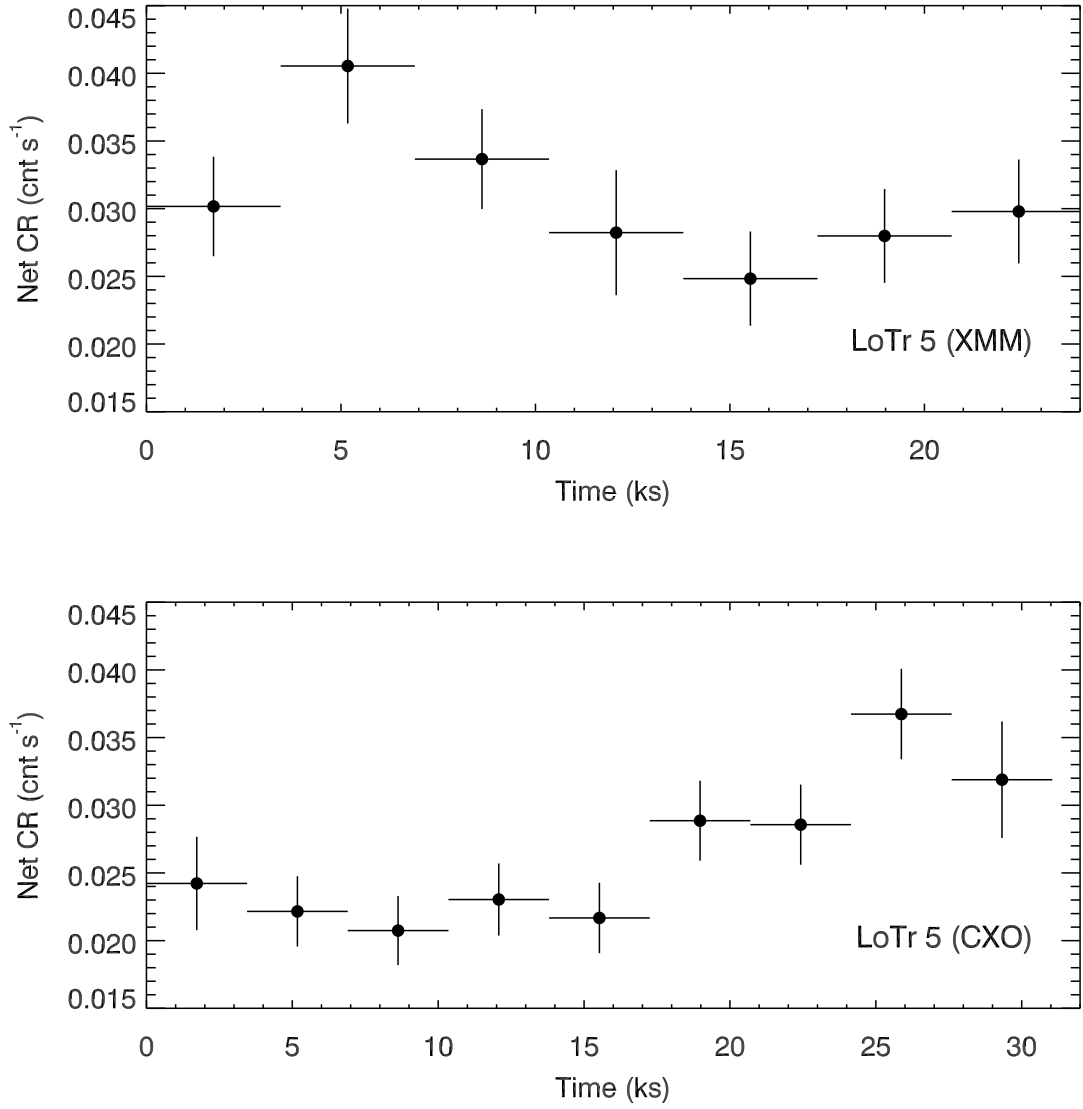}
\caption{\label{lotr5lc} X-ray light curves of LoTr 5 from XMM EPIC pn (upper panel) and the CXO ACIS-S (lower panel) observations.}
\end{figure}

For LoTr 5, if we assume synchronized orbital and rotational periods ($P_{orb} = P_{rot}$), then the binary separation, $A$, is $\sim 16 R_{\sun}$ and the $R_L \sim 7 R_{\sun}$.  
In this scenario, even at the lower limit on the giant's radius \citep[$\sim$7.8 $R_{\sun}$ according to][]{1997A&A...322..511S}, the giant can lose mass to the sdO primary via RLO.
If the orbital period of LoTr 5 is longer than 13 days, the giant is unlikely to be undergoing RLO. 
However, there is evidence of short-term variability (perhaps  rapid flickering) in the optical \citep{1997A&A...322..511S} and longer-timescale variability in the X-ray (Figure \ref{lotr5lc}). 
Both forms of variability could be indicative of accretion on the sdO star, a la CV systems \citep{1992A&A...266..237B}.  
A period of $\sim 0.25$ days was reported by \citet{1993AcA....43..445K} for LoTr 5.   
\citet{1997A&A...322..511S} searched for a similar periodicity but could not find one, leading them to suggest the period found by \citet{1993AcA....43..445K} was an artifact or a time variable phenomenon.    
The origin of this putative 0.25 day period could be a hot spot on an accretion disk around the sdO, or, as 
\citet{1997A&A...322..511S} suggest, a reflection effect off an as-yet undiscovered third component in LoTr 5. 
The X-ray lightcurves in Figure~\ref{lotr5lc} do not clearly support the periodicity proposed by \citet{1993AcA....43..445K}, but we cannot be certain from our data.  
Further observations are required to determine if such a periodicity can be found in the X-ray source in LoTr 5. 

\subsection{X-rays from Internal O-star Wind Shocks}

The shocked winds of O and B stars display log $L_X/L_{\textrm{bol}} \sim -7$ \citep{1994ApJ...421..705C,1981ApJ...250..677C}.
If we attribute the detected X-ray emission to the primary in HFG 1, then log $L_X/L_{\textrm{bol}} \sim -6.1 \pm 0.2$, which is higher than that found in \citet{1994ApJ...421..705C,1981ApJ...250..677C}.
For the primary in HFG 1, a shock velocity of $\sim 1200 \textrm{ km s}^{-1}$ relative to the wind velocity is required to produce the measured 20 MK plasma.
From a radiative driven winds \citep{1988ApJ...335..914O}, this would suggest the wind speed is at least twice the shock velocity, i.e., $>2000 \textrm{ km s}^{-1}$.
As of yet, no such wind has been found in HFG 1; hence, it is unlikely that the X-ray emission arises from shocks in the wind from the primary.
Alternatively, the observed value of $L_{X}/L_{\textrm{bol}}$ could suggest a magnetically active primary, but as we have stated, such a scenario is difficult to reconcile in a star with such a high effective temperature.
Similarly, attributing the X-ray emission from the point source in DS 1 to the primary leads to log $L_X/L_{\textrm{bol}} \sim -6.6 \pm 0.3$, again suggesting a fast shock velocity ($\sim 1000 \textrm{ km s}^{-1}$) or a high level of magnetic activity in a star that is unlikely to support active magnetic fields.

\subsection{X-rays from Coronal Activity Associated with the CSPNe Companions} 

Coronal X-ray luminosity is strongly correlated with rotation for late-type main sequence stars \citep[see review by][]{2009A&ARv..17..309G}.  
This correlation is believed to arise, as in our sun, from dynamo-generated magnetic fields in a differentially rotating late-type star's convective zone.  
For such stars, X-ray luminosity linearly increases with increasing rotation until saturation is reached at a relative X-ray luminosity log $L_X/L_{\textrm{bol}} \sim -3$. 
The origin of this plateau in $L_{X}/L_{\textrm{bol}}$ is unknown, but possible explanations include saturation of the dynamo or limits on the number and size of active areas imposed by the surface area of the star \citep[e.g.][]{2009A&ARv..17..309G}. 
\citet{1993ApJS...86..599D,1993ApJ...413..333D} studied the X-ray emission properties of a large sample of rapidly rotating companions in RS CVn systems, where tidal interactions amongst the companions helps maintain a more rapid rotation than that of single stars at a similar age.  
The X-ray spectra of the coronae of such stars are best described by 2-T thermal plama models comprised of a cool component of a few MK and a hot component of tens of MK \citep{1993ApJ...413..333D}.   
EUV spectra of the coronal emission from nearby active binary systems reveal a continuous emission measure distribution (EMD) that varies with temperature \citep{2003ApJS..145..147S}.
The EMD is explained by \citet{2006ApJ...643..438C} as indicative of a number of unresolved, stable, coronal loops, varying from solar-like loops with modest temperatures ($\sim 2$ MK) and densities ($\sim 10^9$ cm$^{-3}$) to hotter loops ($> 8$ MK) with higher densities ($> 10^{9}$ cm$^{-3}$). 

If we attribute the detected X-ray emission to the main sequence companion in HFG 1, then the activity ratio, log $L_X/L_{\textrm{bol}} \sim -3.0 \pm 0.2$, is at the saturation level.  
Assuming fully ionized coronal plasma, $n_{H} = 0.85 n_{e}$, the emission measure, EM, determined from the X-ray spectral fit is related to the density of coronal plasma by $n_{e} = (EM / 0.85 V)^{1/2}$, where $V$ is the emitting volume.  
The emitting volume can be approximated as a shell on the stellar surface by $V = \frac{4\pi}{3}R_{*}^{3} [ (L+1)^{3} - 1 ]$, where $L$ is the characteristic coronal loop length in stellar radii, $R_{*}$.  
For inactive stars, densities up to $10^{9}\textrm{ cm}^{-3}$ and loop lengths $\sim 0.1 R_{*}$ are typical, while active stars have higher densities, from $10^{10}$ to $10^{11}\textrm{ cm}^{-3}$ and up to $10^{13}\textrm{ cm}^{-3}$ in extreme cases, with generally smaller loop lengths, $L < 0.1 R_{*}$ \citep{2004A&ARv..12...71G}. 
If we take $L = 0.1 R_{*}$, then the lower bound on the density of the hot X-ray gas in HFG 1 is $\sim 10^{10}\textrm{ cm}^{-3}$, consistent with a corona around the companion. 

Applying similar considerations to the X-ray emission detected from DS 1, if we attribute the X-ray emission to the companion, the resulting value of log $L_X/L_{\textrm{bol}} \sim -2.0 \pm 0.3$ is significantly above saturation.
However, the uncertainty in $L_X/L_{\textrm{bol}}$ does not include the poorly constrained uncertainties in the temperature and radius of the companion, which can raise $L_{\textrm{bol}}$ enough to bring $L_X/L_{\textrm{bol}}$ down to the saturation level.   
Nevertheless, a value of $L_X/L_{\textrm{bol}}$ as large as $\sim-2.0$ would not be without precedent.
Indeed, many studies find a large scatter in the $L_X/L_{\textrm{bol}}$ ratio in the saturated region \citep[e.g., the pre-main sequence sample studied by ][]{2004AJ....127.3537S}, with some stars reaching log $L_X/L_{\textrm{bol}} \sim -2.0$. 
The possibility of X-ray flaring activity from DS 1 may also account for the high value of $L_X/L_{\textrm{bol}}$. 
Applying similar coronal assumptions to the X-ray emission detected from DS 1 results in a lower bound on the coronal density of $\sim 10^{10}\textrm{ cm}^{-3}$.

Although the variability of LoTr 5 (in light of the uncertainty in its orbital period) suggests that the X-ray emission detected from LoTr 5 might arise from accretion processes, there is corroborating evidence that supports the notion that the X-rays arise as a consequence of coronal emission from the giant companion.
Good indicators of such coronal activity are the strong Ca II H \& K, Mg II, and H$\alpha$ emission lines, which are likely due to chromospheric activity associated with the giant companion than any accretion-related associated with the sdO primary \citep{1996A&A...307..200J}.
Indeed, the ratio log $L_X/L_{\textrm{bol}} \sim -5$ and projected rotational velocity $v \textrm{ sin } i \sim 67 \textrm{ km s}^{-1}$ \citep{1997A&A...322..511S} of the giant companion in LoTr 5 are consistent with those of rapidly rotating, intermediate mass, G- and K-type giants with coronal activity, both in single systems \citep{2005A&A...444..531G} and binary systems \citep{2007A&A...464.1101G}. 
Such giants do not follow the empirical activity-rotation relation, $L_X \sim 10^{27} (v \textrm{ sin } i)^2$, found by \citet{1981ApJ...248..279P} for main sequence late-type stars and do not appear to reach a saturation level \citep{2007A&A...464.1101G}.   
\citet{2007A&A...464.1101G} interpret the linear relation between the coronal radiative flux density and the average surface magnetic flux density as evidence for increasing magnetic surface flux in response to an increasing angular rotation velocity. 
The X-ray properties of LoTr 5 also agree with those by found by \citet{1993ApJS...86..599D,1993ApJ...413..333D} for RS CVn systems.  
The available evidence indicates that the binary in LoTr 5 may be similar to such systems, as also suggested by \citet{1997A&A...322..511S}. 

\section{Summary}

The X-ray emission sources detected at the binary central stars of three PNe---two suspected post-common envelope binaries, HFG 1 and DS 1, and the binary LoTr 5, whose period is unknown---reveals properties consistent with coronae in late-type, spun up companions.
For HFG 1 and DS 1, the measured ratio of X-ray to bolometric luminosity, $L_X/L_{\textrm{bol}}$, indicates that the main sequence companions are at or beyond the saturation level determined for rapidly rotating late-type stars.  
The X-ray light curves suggest a steady source in HFG 1 and a possibly flaring source in DS 1, which may account for the large relative X-ray luminosity measured for the latter central star, if the companion is the source.
The value of $L_{X}/L_{\textrm{bol}}$ inferred for the companion star in LoTr 5 is similar to that of rapidly-rotating giants and RS CVn systems. 
The X-ray temperatures and emission measures determined from the X-ray spectral fits are furthermore consistent with those found for active, late-type stars.
Although, the X-ray temperatures would also be consistent with those expected from accretion onto the pre-WD, sdO, primary stars, the lack of evidence for Roche lobe overflow from the late-type stars in HFG 1 and DS 1 casts doubt on such an origin for the X-ray emission.
Since the period of the binary in LoTr 5 is unknown, we cannot determine if the late-type giant secondary in LoTr 5 is capable of filling its Roche lobe, and Roche lobe overflow remains a possible source of accreting material in this case. 
However, there is compelling and corroborating evidence that the X-ray emission is due to coronal activity associated with the rapidly rotating giant. 
We conclude that the observed X-ray emission in each of these three binary CSPN systems is most likely due to coronal activity associated with the late-type companions.
Although it is clear that supporting observations are required to detect and characterize any putative companions more directly,
these X-ray observations thereby demonstrate the potential utility of X-ray searches for faint companions in the central stars of PNe.
  
\acknowledgments

This research was supported via award number GO9-0030X to RIT issued by the {\it Chandra} X-ray Observatory Center, which is operated by the Smithsonian Astrophysical Observatory for and on behalf of NASA under contract NAS8--03060, and by NASA Astrophysics Data Analysis grant NNX08AJ65G to RIT.
We are grateful for helpful comments provided by Noam Soker and Jason T. Nordhaus during the preparation of this manuscript.    

{\it Facilities:} \facility{XMM (EPIC), CXO (ACIS)}.

\begin{table*}
\begin{center}
\caption{Summary of Binary Properties$^{a}$\label{binarytable}}
\begin{tabular}{lcccccc}
\tableline
PN & $M_{1},M_{2}$ & $R_{1},R_{2}$ & $T_{\textrm{eff},1},T_{\textrm{eff},2}$ & $L_{\textrm{bol},1},L_{\textrm{bol},2}$ & References \\
   & $(M_{\sun})$ & $(R_{\sun})$ & $(kK)$ & $(L_{\sun})$ &  \\
\tableline
HFG 1 & 0.57,1.09 & 0.19,1.30 & 83,5.4 & 1500,1.3 & 1 \\
HFG 1 & 0.63,0.41 & ...,1.15 & ...,5.3 & ...,0.9 & 2 \\
DS 1  & 0.63,0.23 & 0.157,0.402 & 77,3.4 & 776,0.02 & 3 \\
LoTr 5 & 0.6,1.1 & 0.05,8-12 & 185,5.3 & 2600,100 & 4,5,6 \\
\tableline\end{tabular}
\tablenotetext{a}{The stellar properties of the primary and secondary components are labeled by the 1 and 2 subscripts, respectively.} 
\tablenotetext{b}{References: 1-Shimanski et al. 2004, 2-Exter et al. (2003) solution for $K_{p} = 49$ km s$^{-1}$ and $i = 29^{\circ}$, 3-Hilditch et al. (1996), 4-Feibelman \& Kaler (1983), 5-Jasniewicz et al. (1996), 6-Graham et al. (2004)}
\end{center}
\end{table*}

\begin{table*}
\begin{center}
\caption{Summary of X-ray Observations \label{obstable}}
\begin{tabular}{lccccc}
\tableline
Object & Date & ObsID & Chip & t$_{exp}$ & Net CR \\
       &      &       &      &   (ks) & (cnt ks$^{-1}$) \\
\tableline
HFG 1   & 2008 Dec 11     & 9954 & ACIS-S3       & 11.3 & 12.4 \\
DS 1   & 2009 Jul 19     & 9953 & ACIS-S3       & 23.8 & 2.30 \\
LoTr 5 & 2002 Jun 06     & 12850201 & EPIC pn   & 20.9 & 34.7 \\
       & ...             & ...      & EPIC MOS1 & 25.7 & 11.3 \\
       & ...             & ...      & EPIC MOS2 & 25.7 & 8.78 \\
LoTr 5 & 2002 Dec 04     & 3212 & ACIS-S4       & 27.7 & 26.1 \\
\tableline
\end{tabular}
\end{center}
\end{table*}

\begin{landscape}
\begin{table*}
\begin{center}
\caption{Spectral Fitting of the X-ray Emission from Binary CSPNe\label{specfits}}
\footnotesize
\begin{tabular}{lccccccccccc}
\tableline
PNe & $\chi^{2}$ (d.o.f.$^{a}$) & log $N_H$ & kT & $T_X$ & log norm.$^{b}$ & log EM$^{b}$ & log $F_{X,\textrm{obs}}$ & log $F_{X,\textrm{unabs}}$ & log $L_{X}$ \\
    &   &  (cm$^{-2}$) &  (keV) & (MK) & (cm$^{-5}$) & (cm$^{-3}$) & (erg cm$^{-2}$ s$^{-1}$) & (erg cm$^{-2}$ s$^{-1}$) & (erg s$^{-1}$) \\
\tableline 
HFG1 & 0.79 (510) & 21.4 & 0.5$_{-0.1}^{+0.1}$ & 5.8 & -4.5$_{-0.1}^{+0.1}$ & 53.10 & -13.57 & -13.13 & 30.51 \\
 &  &  & 1.8$_{-0.5}^{+1.1}$ & 21.3 & -4.5$_{-0.2}^{+0.2}$ & 53.11 & -13.56 & -13.39 & 30.25 \\
DS 1 & 0.49 (510) & 20.9 & 0.3$_{-0.1}^{+0.2}$ & 3.1 & -5.7$_{-0.5}^{+0.2}$ & 52.07 & -14.65 & -14.45 & 29.32 \\
 &  &  & 1.2$_{-0.2}^{+0.4}$ & 14.5 & -5.3$_{-0.2}^{+0.2}$ & 52.47 & -14.14 & -14.06 & 29.71 \\
LoTr 5 & 1.34 (34) & 0.00 & 0.65$_{-0.05}^{+0.05}$ & 7.6 & -4.93$_{-0.08}^{+0.05}$ & 52.54 & -13.54 & -13.54 & 29.93 \\
(XMM)  &  &  & 2.27$_{-0.35}^{+0.51}$ & 26.3 & -4.37$_{-0.05}^{+0.06}$ & 53.11 & -13.22 & -13.22 & 30.25 \\
LoTr 5 & 0.50 (23) & 0.00 & 0.64$_{-0.07}^{+0.07}$ & 7.5 & -4.71$_{-0.09}^{+0.10}$ & 52.77 & -13.30 & -13.30 & 30.18 \\
(CXO)  &  &  & 3.49$_{-0.64}^{+0.83}$ & 40.5 & -3.93$_{-0.05}^{+0.04}$ & 53.55 & -12.74 & -12.74 & 30.73 \\
\tableline
\end{tabular}
\tablenotetext{a}{When fitting the unbinned spectra of HFG 1 and DS 1, the empty bins are used as constraints, hence the inflated number of degrees of freedom (d.o.f.).} 
\tablenotetext{b}{Emission Measure (EM) is determined from the normalization value of the \textit{mekal} model according to:
\begin{equation} 
\textrm{EM} = \int n_e n_H dV = 1.2 \times 10^{58} \textrm{ cm}^{-3} \left[ \textrm{ norm. } \times \left( \frac{ D }{ 1 \textrm{ kpc}} \right)^2 \right] 
\end{equation}
}
\normalsize
\end{center}
\end{table*}
\end{landscape}


\end{document}